\documentclass[prl,twocolumn,showpacs,preprintnumbers,cite]{revtex4}

\usepackage{graphicx}
\usepackage{dcolumn}
\usepackage{bm}

\usepackage{amssymb}
\usepackage{amsmath}

\newcommand{\pos}{PrOs$_{4}$Sb$_{12}$}

\newcommand{\msr}{$\mu$SR}

\begin{document}

\preprint{}

\title{Time-Reversal Symmetry-Breaking Superconductivity in Heavy Fermion PrOs$_4$Sb$_{12}$ detected by Muon Spin Relaxation}

\author{Y.~Aoki}
\author{A.~Tsuchiya}
\author{T.~Kanayama}
\author{S.R.~Saha}
\author{H.~Sugawara}
\author{H.~Sato}
\affiliation{Department of Physics, Tokyo Metropolitan University, Hachioji, Tokyo 192-0397, Japan}%
\author{W. Higemoto}
\author{A. Koda}
\author{K. Ohishi}
\author{K. Nishiyama}
\author{R. Kadono$^*$}
\affiliation{Institute of Materials Structure Science, High Energy Accelerator Research Organization (KEK), Ibaraki 305-0801, Japan}%

\date{\today}

\begin{abstract}
We report on muon spin relaxation measurements of the $4f^2$-based heavy-fermion superconductor filled-skutterudite PrOs$_4$Sb$_{12}$.
The results reveal the spontaneous appearance of static internal magnetic fields below the superconducting transition temperature, providing unambiguous evidence for the breaking of time-reversal symmetry in the superconducting state.
A discussion is made on which of the spin or orbital component of Cooper pairs carries a nonzero momentum.

\end{abstract}



\maketitle
Many unconventional superconducting (SC) states, as in Ce- and U-based heavy-fermion (HF) compounds or high-$T_c$ cuprates, appear in close proximity to magnetic instabilities when a certain parameter (pressure, atomic doping, or oxygen content) is varied~\cite{Mathur98,UGe2,PhysToday}.
This fact strongly suggests that the attractive interactions binding electrons into Cooper pairs are mediated by magnetic moment fluctuations.
As another possible pairing glue, fluctuations of quadrupole moments---distorted shapes of the electronic clouds of ions, are theoretically considered to be possible~\cite{Co98,No94}.
An interesting question to be addressed is what is the nature of superconductivity in such a system.
For this study, Pr-based compounds with a $4f^2$ configuration are likely candidates, since nonmagnetic but quadrupolar active low-energy levels can be realized due to the crystalline-electric-field (CEF) effect; in $5f$ systems, CEF levels are less clear due to the tendency to be itinerant.

One promising candidate material for this study is the recently found superconductor \pos\ \cite{BauerPRB}, which is to date the only known Pr-based HF superconductor, with a superconducting transition temperature of 1.82 K (hereafter referred to as $T_{c1}$).
The estimated electronic specific heat coefficient $\gamma=350-700$ mJ/K$^2$mol~\cite{BauerPRB,MapleOrb} and the enhanced cyclotron-effective masses~\cite{SuDHvA} reflect the existence of strong electron correlations.
Specific heat ($C$), magnetic susceptibility ($\chi$), and inelastic neutron studies provide evidence that \pos\ has a nonmagnetic ground state and a magnetic triplet excited state separated by $\Delta E_{\rm CEF}/k_B=8 $ K~\cite{MapleOrb,AokiJPSJ}, which is 5 times larger than $T_{c1}$.
In the temperature-versus-magnetic-field ($T$-vs-$H$) phase diagram, a field-induced ordered phase ($\mu_0 H \gtrsim4 $ T)~\cite{AokiJPSJ} appears close to the superconducting phase (the upper critical field $\mu_0 H_{c2}=2.2 $ T).
It was recently proven to be an antiferro-quadrupolar ordered phase by elastic neutron scattering measurements~\cite{KohgiJPSJ}.
This fact strongly indicates that quantum quadrupole fluctuations of the Pr ions play an important role in realizing the HF superconductivity in PrOs$_{4}$Sb$_{12}$, considering that the $T$-vs-$H$ phase diagram is analogous to those for the HF and cuprate systems, where a magnetically ordered phase exists close to the SC phase in the $T$-vs-pressure, -atomic-doping, or -oxygen-content phase diagram.
This scenario is further supported by the enhanced $T_{c1}$ compared to $T_c=0.74 $ K for a $4f^0$ reference compound LaOs$_4$Sb$_{12}$~\cite{SuDHvA,NQR}. 

The remarkable unconventional SC properties described below suggest non-$s$-wave pairing in \pos.\ 
Thermal conductivity measurements in magnetic fields rotated relative to the crystal axes indicate the presence of two distinct superconducting phases with different symmetries with point nodes~\cite{Izawa}.
The specific heat also suggests an additional phase transition at $T_{c2}=1.65-1.75$ K $<T_{c1}$~\cite{MapleOrb,Vollmer}.
Sb-nuclear-quadrupole-resonance (NQR) spin-lattice-relaxation rate $1/T_1$ exhibits no Hebel-Slichter peak just below $T_{c1}$~\cite{NQR}.
In contrast, exponential $T$ dependences of $1/T_1$ in the lower $T$ region~\cite{NQR} and of the magnetic penetration depth determined by a transverse-field muon-spin-rotation study~\cite{MacPRL} indicate, however, an isotropic energy gap.
These behaviors, making it difficult to construct a simple picture for the SC state, suggest that a novel type of HF superconductivity is realized in this compound.

One important aspect for characterizing a SC state is whether time-reversal symmetry (TRS) is broken or not~\cite{SigUeda}.
In a state with broken TRS, the magnetic moments of Cooper pairs are nonzero and align locally, and thereby a spontaneous but extremely small internal magnetic field can appear. 
To detect such fields, zero-field (ZF) muon spin relaxation (\msr) is the most powerful method, because of its high sensitivity, as small as 10 $\mu$T~\cite{Schenck,Amato}.
TRS-broken SC states are quite rare, since an unambiguous observation of such fields has been made only in Sr$_2$RuO$_4$~\cite{nSRO}.
In the low-$T$ phase of Th-doped UBe$_{13}$, the possibility of a magnetic phase coexisting with the SC state cannot be ruled out~\cite{Heff} and, on an observation in UPt$_3$~\cite{LukeUPt3}, doubt has been cast later~\cite{DaUPt3,HigeUPt3}.

In this Letter, we report on ZF-\msr\ measurements for \pos.
The results unambiguously reveal the spontaneous appearance of an internal magnetic field in the SC state, providing clear evidence for broken TRS.
This is the first observation of such superconductivity in {\it undoped} intermetallic compounds.

Conventional $\mu$SR measurements were carried out down to 20 mK using a top-loading dilution refrigerator at the $\pi$A-port of the Meson Science Laboratory, KEK-MSL, Japan.
Small single crystals of PrOs$_4$Sb$_{12}$ with sizes of $\lesssim 1$ mm grown by the Sb-flux method~\cite{SuDHvA} using raw materials of 4N(99.99\% pure)-Pr, 3N-Os, and 6N-Sb were used.
Clear de Haas-van Alphen (dHvA) oscillations observed in one of the crystals~\cite{SuDHvA} are indicative of their high quality.
The single crystals with random orientation were glued onto a 4N-silver holder covering $\sim 25 $ mm$\phi$.
The silver holder was held in a vacuum space of the refrigerator by screwing it to a copper rod thermally connected to the mixing chamber.
Stray fields at the sample position were cancelled to within 1 $\mu$T in all directions by using three pairs of correction coils and a flux-gate magnetometer.
We implanted spin-polarized positive muons ($\mu^+$) into the sample and the time evolution of the $\mu^+$ spin polarization was measured using positron detectors positioned parallel and antiparallel to the initial polarization direction.

\begin{figure} 
\includegraphics[scale=0.5]{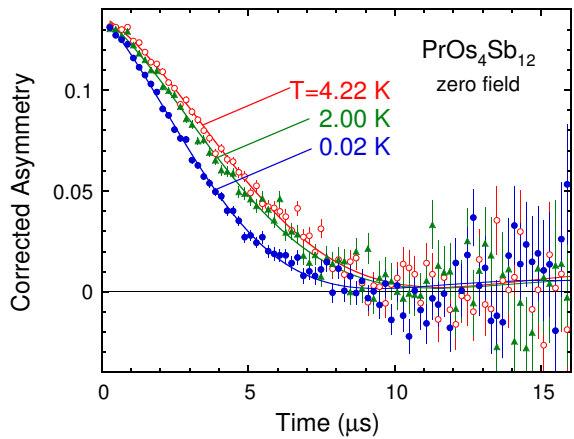}
\caption{Zero-field $\mu$SR spectra in \pos\ measured across the superconducting transition temperature $T_{c1}=1.82 $ K. 
The corrected asymmetry was obtained by subtracting the temperature-independent contribution from the silver plate (0.057).
The curves are fits to the model I given by Eq.~(1). The relaxation rate becomes stronger below $T_{c1}$.}
\label{fig:spectra}
\end{figure}

\begin{figure}
\includegraphics[scale=0.45]{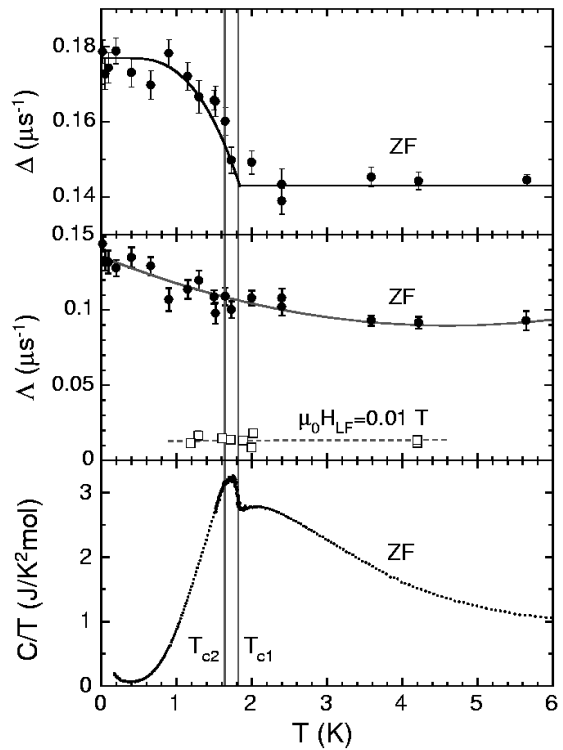}  
\caption{Temperature dependences of the width of the internal field distribution $\Delta$, the relaxation rate $\Lambda$ in zero-field (ZF) and in $\mu_0 H_{\rm LF}=0.01 $ T, and the specific heat divided by temperature $C/T$. The vertical lines indicate two consecutive transitions at $T_{c1}$ and $T_{c2}$ determined from the $C/T$ data.}
\label{fig:Tdep}
\end{figure}

The time evolution of the ZF muon spin polarization is shown in Fig.~\ref{fig:spectra} for $T=4.2, 2.0,$ and 0.02 K, where a nonrelaxing background signal originating from $\mu^+$ stopping mainly in the silver holder had already been subtracted.
No signature of precession is visible, ruling out the presence of a sufficiently large internal magnetic field as seen in magnetically ordered compounds.
It is obvious from Fig.~\ref{fig:spectra} that the relaxation becomes stronger with decreasing $T$.
We tried several functional forms to reproduce the ZF spectra, and excellent fits to the data were obtained using two models, I and II.
The fitting using model I,
\begin{gather}
P_\mu = \exp(-\Lambda t) G_z^{\rm KT}(\Delta, t) , \notag \\ 
G_z^{\rm KT}(\Delta, t) =  \frac{1}{3}+\frac{2}{3}(1-\Delta^2 t^2) \exp(-\frac{1}{2}\Delta^2 t^2) , \label{eq:modelI}
\end{gather}
which is commonly used~\cite{nSRO,LukeUPt3,MacPRL}, is shown by the solid curves in Fig.~\ref{fig:spectra}.
The Kubo-Toyabe (KT) function $G_z^{\rm KT}$ is attributed to the muon-spin relaxation due to static randomly-oriented local fields at the muon site (caused by nearby nuclei in the normal state), and $(\Delta/\gamma_\mu)^2$ ($\gamma_\mu=8.516 \times 10^8 $ s$^{-1}$T$^{-1}$: the $\mu^+$ gyromagnetic ratio) represents the second moment of the field distribution.
The necessity of $\exp(-\Lambda t)$ indicates the presence of an additional relaxation process.
The best-fit values of $\Delta$ and $\Lambda$ are shown in Fig.~\ref{fig:Tdep} along with the specific heat divided by temperature $C/T$ measured using a single crystal from the same batch.
Around the SC transition, $C/T$ shows two types of anomalies, i.e., a sharp jump at $T_{c1}=1.82 $ K and a kink structure at $T_{c2}=1.64 $ K, in a slightly different way from the reported data~\cite{MapleOrb,Vollmer}.

It is remarkable that $\Delta$ shows a significant increase with an onset temperature of around $T_{c1}$, indicating the appearance of a spontaneous internal field correlated with the superconductivity; this behavior is in marked contrast with that of $\Lambda$, which increases gradually with decreasing $T$ across $T_{c1}$ and $T_{c2}$ without showing any distinct anomalies.
This observation provides unambiguous evidence that TRS is broken in the SC state of \pos.
Note that the observed increase in $\Delta$ is smaller than the experimental uncertainty in the previous ZF-\msr\ measurement~\cite{MacPRL}.
Provided that the increase is of electronic origin, the electronic ($\Delta_e$) and nuclear ($\Delta_n$) contributions to $\Delta$ must be uncorrelated and thereby their respective contributions add in quadrature: $\Delta_e(T)^2+\Delta_n^2=\Delta(T)^2$.
From Fig.~\ref{fig:Tdep}, $\Delta_n/\gamma_\mu=1.7 \times 10^{-4} $ T and $\Delta_e(T \to 0)/\gamma_\mu=1.2 \times 10^{-4} $ T are derived.
The temperature dependence of $\Delta_e$ below $T_{c1}$ agrees with that of the BCS order parameter $\Delta_{BCS}(T)$, as is evident in Fig.~\ref{fig:Tdep}, supporting our interpretation.

The equally probable model II,
\begin{equation}
 \label{eq:modelII} P_\mu=\exp(-\Lambda t) [\frac{1}{3}+\frac{2}{3}\cos(\omega_e t)] G_z^{\rm KT}(\Delta_n, t) ,
\end{equation}
also fits the data, yielding almost the same $\chi^2$ statistics as model I; i.e., the two models are experimentally undistinguishable.
This fact points to an alternative possibility that the spontaneous fields have almost the same magnitude ($\mu_0 H_{e}=\omega_e/\gamma_\mu$) for all $\mu^+$ sites in the sample. 
Extracted $H_{e}(T)$ data show a similar temperature dependence as $\Delta_e(T)$ and approach $1.9 \times 10^{-4} $ T as $T \to 0$ (not shown).

As the origin of the increased relaxation in the SC state, the following spurious effects can be ruled out.
Any inhomogeneous distribution of extremely small residual fields ($< 1 \mu$T $\ll \Delta_e(T \to 0)/\gamma_\mu \ll \mu_0 H_{c1} \sim 5 \times 10^{-3} $ T~\cite{MacPRL}) caused by the superconducting transition cannot explain the observed anomaly.
A null experiment confirmed that the 4N-silver holder does not exhibit any noticeable anomalies.
The absence of such an anomaly in the previous measurements for UPt$_3$~\cite{HigeUPt3}, CeIrIn$_5$, and CeCoIn$_5$~\cite{Hige115}, performed under the same conditions, also demonstrates that the observed anomaly is uniquely attributed to the PrOs$_4$Sb$_{12}$ specimen.

Any lattice effects are also ruled out for the following reasons:
The observed nuclear dipolar field of $1.7 \times 10^{-4} $ T is close to $2.0 \times 10^{-4} $ T calculated for a most probable $\mu^+$ stopping site (1/2, 0, 0.15) ($12e$ in Wyckoff notation, space group $Im \overline{3}$), which is one of the interstitial sites of Sb-icosahedron cages and is surrounded by six nearest Sb ions.
This site is the same as the one expected for an isostructural PrFe$_4$P$_{12}$, as suggested from transverse-field \msr\ measurements~\cite{PFP}. 
A simple calculation shows that a lattice shrinkage of $\sim 7$\% would be necessary to explain the observed increase in $\Delta$ when $\Delta_e=0$.
This scenario contradicts a thermal expansion measurement~\cite{Oeschlar}.
Furthermore, the electric field gradient at the Sb nuclei~\cite{NQR} does not show any corresponding anomalies across the SC transition.

The spontaneous internal field has two types of possible sources depending on the spin and/or orbital parts of Cooper pairs having nonzero values: (i) (for both nonzero spin and orbital moments) spontaneous undumped supercurrents induced in the vicinity of impurities, surfaces, and/or domain walls between the degenerate SC phases, where the order parameter has spatial inhomogeneities~\cite{SigUeda,Choi,Mineev}, and
(ii) (for nonzero spin moments) a finite hyperfine field induced at the $\mu^+$ sites.

An example of the source (i) is found in Sr$_2$RuO$_4$~\cite{nSRO}, where $\Lambda$ exhibits a spontaneous increase below $T_c$.
This can be understood by considering that a dilute distribution of the sources in the sample would result in a Lorentzian-type field distribution~\cite{Uemura}.
Although it is possible that the present observation in \pos\ is also due to the same type of sources, the spontaneous increase appears in $\Delta_e$ instead of $\Lambda$.
This fact suggests that the field distribution is rather of Gaussian type and the second moment has a finite value.
The mean-free path $\ell \sim 2000 $ \AA \ estimated from the Dingle temperature (proportional to the electron scattering rate) in the dHvA experiments~\cite{SuDHvA}, which is much larger than the SC coherence length $\xi_0=120 $ \AA, suggests that the sample is in a clean limit.
If the scattering centers determining the Dingle temperature are the main sources of the spontaneous fields, the average nearest-neighbor distance ($ \sim \ell$) would be comparable to the magnetic penetration depth $\lambda=3400 $ \AA~\cite{MacPRL}.
This could explain why the field distribution is of Gaussian type.
Although quantitative comparisons with the models~\cite{Choi,Mineev} are hampered by a lack of detailed information on the sources, it is certain that the field at the center of the sources should be much larger than $\Delta_e/\gamma_\mu$.

The field source (ii) is possible only if the SC pairing symmetry is not only of spin triplet (odd parity), but also of "nonunitary" type (analogous to the A$_1$ phase of superfluid $^3$He~\cite{3He}).
In such a state, the three-dimensional vector function {\boldmath $d(k)$}, representing the wave-vector dependent SC order parameter~\cite{SigUeda}, is complex and a nonzero spin moment given by {\boldmath $d(k)$}$\times${\boldmath $d(k)$}$^*$ could produce a finite hyperfine field on $\mu^+$.
Note that no experimental identification of such nonunitary superconductivity has ever been reported.
If this is the case in \pos, $\mu_0 H_{e}$ corresponds to the "intrinsic" spontaneous field at $\mu^+$ sites, which is irrelevant to the existence of impurities.
The magnitude of the field can be theoretically estimated to be $\sim 7 \times 10^{-4} \eta  $ T~\cite{Machida}, where $\eta$ denotes the degree of the electron-hole asymmetry of the conduction band at the Fermi level and is of the order of $T_{c}/T_F$ ($T_F$ is the Fermi temperature or the Kondo coherence temperature of the heavy quasiparticle band).
A simple estimation of $\sim 1\times 10^{-4} $ T using $T_F=10 $ K, which is given by the temperature dependence of $T_1$~\cite{NQR}, is in good agreement with the present observation.
A unique feature of the nonunitary state is that the spin-up and spin-down Cooper pairs have different sizes of the excitation gap~\cite{SigUeda,Machida}.
The smaller gap branch with $|${\boldmath $d(k)$}$|^2-|${\boldmath $d(k)$}$\times${\boldmath $d(k)$}$^*| \neq 0$ might explain apparent residual contributions appearing at $T \ll T_{c1}$ in some physical quantities~\cite{NQR,Izawa}.

\begin{figure}
\includegraphics[scale=0.5]{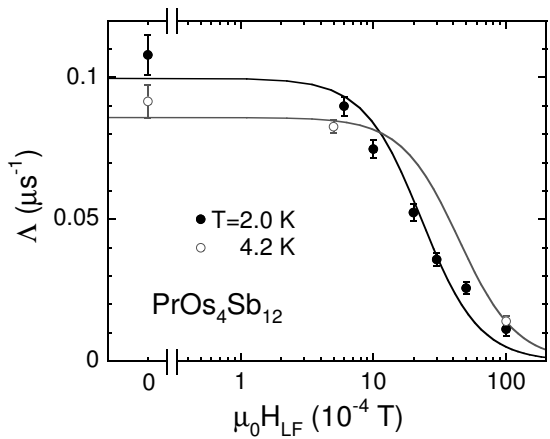}  
\caption{Longitudinal field $H_{\rm LF}$ dependence of the relaxation rate $\Lambda$. The curves are fits to the model given by Eq.~(3).}
\label{fig:LF}
\end{figure}

We performed measurements in weak longitudinal fields (LF) ($\mu_0 H_{\rm LF}$), parallel to the initial $\mu^+$ polarization, and confirmed that the spontaneous internal fields are static on the microsecond time scale.
In the mixed state after field cooling in $\mu_0 H_{\rm LF} = 100 \times 10^{-4} $ T, the relaxation due to the spontaneous fields is no longer visible (already decoupled), and excellent fits to the spectra are obtained using only the $\exp (-\Lambda t)$ term.
The temperature dependence of $\Lambda$ in this field (see Fig.~\ref{fig:Tdep}) exhibits no distinct anomaly through the SC transition.

In zero field, $\Lambda$ shows a gradual increase with decreasing temperature even in the normal state, in contrast to other HF superconductors~\cite{LukeUPt3,HigeUPt3,Heff}.
The $H_{\rm LF}$ dependence of $\Lambda$ shown in Fig.~\ref{fig:LF} can be nicely fitted by the Redfield model:
\begin{equation}
\label{eq:RF} \Lambda=2 (\gamma_\mu \mu_0 H_{\rm loc})^2 \tau_c/[1+(\gamma_\mu \mu_0 H_{\rm LF} \tau_c)^2 ],
\end{equation}
indicating that this relaxation process is due to fluctuating local fields characterized by the magnitude $\mu_0 H_{loc}=3.7\ (4.7) \times 10^{-4} $ T and the correlation time $\tau_c=0.51\ (0.27)\ \mu$s for 2.0 (4.2) K.
We expect that the relaxation process represented by $\Lambda$ may reflect the $4f$-electron dynamics associated with the gap feature given by $\Delta E_{\rm CEF}$.

In conclusion, the ZF-\msr\ measurements in \pos\ have revealed an appreciable increase in the internal magnetic fields below around the onset of superconductivity ($T_{c1}=1.82 $ K).
This provides clear evidence for the broken TRS in the SC state, which will help us to narrow down the number of possibilities for the symmetry of the SC order parameter.
From the present results, however, a definite conclusion cannot be drawn regarding which part of the Cooper pairs, spin or orbital (or both), carries a nonzero magnetic moment.
The broken TRS indicates that the SC state belongs to a degenerate representation, which has internal degrees of freedom.
This is in line with the possible existence of multiple superconducting phases suggested by the specific heat and thermal conductivity studies.
Considering the normal state properties, where quadrupolar degrees of freedom of Pr ions play a major role, the present results point to an exotic conjecture that the {\it magnetic} and heavy Cooper pairs in \pos\ are formed by {\it nonmagnetic} interactions mediated by quantum quadrupole fluctuations.
Although this needs to be substantiated further by other means, \pos\ provides us with a new, and apparently unique, example of superconductivity in the field of strongly correlated electron systems.

We thank J. Goryo, K. Izawa, M. Koga, K. Machida, Y. Matsuda, K. Miyake and M. Sigrist for valuable discussions.
We are grateful to T. Namiki for his assistance.
This work was supported by a Grant-in-Aid for Scientific Research from MEXT of Japan and the Inamori Grant Program.

$^*$ Also at the Graduate University for Advanced Studies (SOKENDAI).

\end{document}